\documentclass[pre,twocolumn,showpacs,preprintnumbers]{revtex4-1}
\usepackage{amsmath,amssymb,latexsym} 
\usepackage{color}

\usepackage{graphicx} 
\usepackage{natbib}
\usepackage{epstopdf}
\usepackage{color}
\usepackage{makecell}
\usepackage[table]{xcolor}
\usepackage{tabularx}
\usepackage{ragged2e}

\begin{document}
\clearpage



\title{A calculation of the Deuterium Hugoniot using the\\
classical-map hypernetted-chain (CHNC) approach.}

\author{M. W. C. Dharma-wardana 
\email[Email address:\ ]{chandre.dharma-wardana@nrc-cnrc.gc.ca}}
\affiliation{
National Research Council of Canada, Ottawa, Canada, K1A 0R6
}
%
\begin{abstract}
The Hugoniot for Deuterium is calculated using the classical-map
hyper-netted-chain (CHNC) approach using several models of the
effective temperature that may be assigned to the electron-ion
interaction. This effective temperature embodies the exchange-correlation
and kinetic energy functional that is assigned to the
electron-ion interaction. 
 Deuterium pair distribution functions (calculated using the
 neutral-pseudo atom method)
showing the formation
of molecular pre-peaks are displayed to clarify the
soft-turning over of the hugoniot in the pressure
range of 0.2-0.6 Megabars.
This contribution updates a previous
CHNC calculation of the deuterium hugoniot given
in Phys. Rev. B, {\bf 66}, 014110 (2002).
\end{abstract}
\maketitle

\section{Calculations of the Deuterium Hugoniot using a classical
 representation of the electrons.}

The major problem in simulations of warm dense matter is the quantum nature of
electrons and the representation of physical variables by operators defined in
function space. The higher the temperature, the larger is the number of excited
states needed to properly represent this function space. The large basis sets
needed make the usual density-functional theory (DFT) and molecular dynamics
(MD) method prohibitive for temperatures $T/E_F$ exceeding about 0.5, where
$E_F$ is  the Fermi energy. At low temperatures, the mean free paths 
$\lambda_{ mfp}$ of electrons become very large, small-$k$ effects become
relevant, and large MD simulation cells with the number of ions $N$ exceeding
thousands~\cite{Pozzo11} become necessary, making the method  impractical
or inaccurate for many physical properties.
 
However, DFT offers the possibility of completely describing at least the
static properties, e.g., the equation of state (EOS) of quantum systems using
only the one-body electron density $n(r)$, {\it without appeal} to
wavefunctions. Hence, a classical representation of electrons which interact 
via an equivalent effective potential can be legitimately
anticipated~\cite{chnc1} if the effective potentials accurately
predict the DFT $n(r)$. Thus the Coulomb interaction $V^{ee}_{cou}(r)$
inclusive of quantum diffraction corrections, and an additional Pauli-exclusion
interaction $P^{ee}(r,\zeta)$  chosen to exactly recover  the Fermi hole in the
electron-electron pair distribution function (PDF) has to be constructed. That is,
the e-e  PDF, $g_{ee}(r,\zeta)$   which depends on the spin polarization
$\zeta$   needs to  be mapped to an effective classical potential
$V^{ee}_{cls}(r)=V^{ee}_{cou}(r)+P^{ee}(r)$ even at $T=0$. However, such an
equivalent Coulomb system at the physical temperature $T$ will also have an
effective classical temperature $T^{ee}_{cls}$ which takes account of the
non-zero kinetic and correlation effects which manifest even at $T=0$. In
effect, $T^{ee}_{cls}$ is an approximate {\it classical} solution to the long
standing problem of the representation for the kinetic energy functional of
DFT.

Thus the self-consistent construction of $V_{cls}(r,\zeta,\bar{n})$, and
$T^{ee}_{cls}(\zeta, \bar{n})$ for a given mean electron density $\bar{n}$
and temperature $T$ to accurately recover the one-body density $n(r)$
and the associated PDFs using the classical hyper-netted-chain equation
 constitutes the classical map of the quantum electrons of spin polarization
 $\zeta$, denoted by CHNC. If molecular dynamics is used instead of the HNC,
for inverting or calculating the PDFs,  we refer to it as  CMMD.

Pair-distribution functions $g_{ee}(r,\zeta)$ using such a map $V^{e}_{cls},
T^{ee}_{cls}$ can be generated via MD or even more cheaply via the hyper-netted
chain (HNC) equation or its modified form known as MHNC containing an
appropriate bridge-diagram correction. 
Dharma-wardana and  Perrot have constructed such maps for 2-D
and 3-D electrons at zero and finite $T$~\cite{chnc1,chnc2}, and shown
accurate agreement of the PDFs with quantum Monte Carlo (QMC) PDFs available at
$T=0$, even for very high coupling, while Dufty {\it et al.}~\cite{SandipDufty13}
 have also constructed similar classical maps. Dufty {\it et al.} have
used additional constraints in constructing potentials rather than
using the minimum necessary conditions imposed by DFT and HNC or MD invertible
maps. Their methods also link with the older work on effective
potentials derived from Slater sums. 

In the classical map proposed by Dharma-wardana and Perrot
 we use an effective temperature $T^{ee}_{cls}$
which is chosen so that the equivalent 
classical Coulomb fluid has the correct exchange-correlation
(XC) energy at $T=0$. This is found to be sufficiently accurately given by
the form
\begin{equation}
T^{ee}=\{T_q^2+T^2\}^{1/2}.
\end{equation} 
The ``quantum temperature'' $T_q$ is of the order of $E_F$;  its
form proposed in Ref.~\cite{chnc1} 
 is a simple function of the electron Wigner-Seitz
 radius $r_s=\{3/(4\pi\bar{n}\}^{1/3}$. The use of the
actual form and the Pauli potential are crucial to the
accuracy of the map.
This   classical map given  by Dharma-wardana and
Perrot in 2000 was found to be very accurate when   results
for finite-$T$ QMC exchange-correlation energies
~\cite{BrownXCT13,Hungary16,TrickeyXC14} and finite-$T$
$g_{ee}(r,\zeta)$  became
 available a decade later.

\section{Classical map for electron-ion mixtures.}
\begin{figure}[t]
\includegraphics[width=\columnwidth]{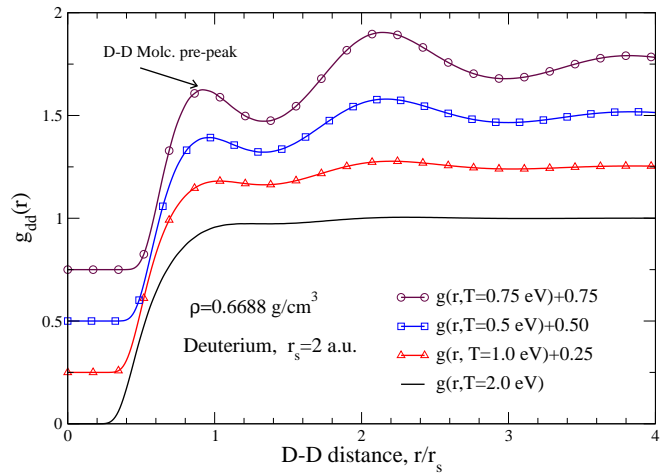}
\caption{(Color online) The onset of transient D-D molecular bonding is 
signaled by the appearance of a pre-peak in the deuterium-deuterium PDF
as the temperature is lowered from 2 eV to 0.3 eV. These calculations use
 the Neutral
pseudo-atom NPA) model which is a parameter-free DFT model~\cite{Hungary16}.
Note that at $r_s$ = 2 there are no 1$s$-bound states as yet, and
the D-D molecules are not stable molecules. As $T$ is lowered, the pre-peak
position converges towards the D-D covalent bond distance. The emergence of the
transient D$_2$ molecules from the plasma state
 does not involve a phase transition as in carbon~\cite{cdw-carbon16}.}
\label{gHH.fig}
\end{figure}

Once the electrons are mapped to an effective classical system, they can now be
used with ions (e.g., protons, carbon nuclei etc.) to study the properties
 of electron-ion systems
without the heavy computational burden of purely quantum approaches. However,
in including the ions, the electron-ion interaction itself needs to be mapped
to a classical form correctly so as to include (i) quantum effects in the
effective electron-ion interaction; (ii) quantum effects in the kinetic energy
functional applicable to the effective temperature $T^{ei}_{cls}$ of the
electron-ion interaction. The ions themselves can usually be taken to be
classical and  pose no problems, with the ion temperature $T_i=T$.  In dealing
with $V^{ei}_{cls}$ we need to consider  (a) that the quantum electron-ion interactions
produce a bound-electron spectrum requiring the use of a
pseudo-potential with a core radius $r_c$ for $V_{cls}^{ei}$, (b) the fact that transient
 as well
as persistent molecular forms, e.g., H-H, C-H and C-C are
formed~\cite{cdw-carbon16}.  Examples of H-H or equivalently D-D PDFs showing
 pre-peaks in 
$g_{dd}(r)$ due the formation of transient D-D bonds, calculated using the
neutral-pseudo-atom (NPA) approach~\cite{cdw-cpp15}, are shown in
Fig.~\ref{gHH.fig}. Unlike many average-atom models, the NPA approach
is capable of providing an average description of all transiently bonded
forms that may be found in the fluid, as described in Ref.~\cite{cdw-carbon16}.
The CHNC currently does a less satisfactory job of picking up such effects,
as this depends on the specification of $T^{ei}$.

The CHNC is developed within the DFT conceptual framework where one-body densities
and XC-functionals rather than wavefunctions   determine the physics of the system.
We recall~\cite{ilciacco93} that an ion-electron system with one-body densities
$n(r)$ for electrons and $\rho(r)$ for ions has  a free energy of the
form $F([n],[\rho])$ and hence the XC-functionals and Kohn-Sham
potentials need to be calculated via functional derivatives with respect to
$n(r)$ and also  $\rho(r)$, and not just with respect to $n(r)$.
Obtaining  XC-potentials needs taking  further functional derivatives. Unlike in 
conventional DFT, the ions are not treated as merely providing a static
``external potential'' but enter directly into the theory. This leads to two 
DFT equations (one for the ions, and one for the electrons),
and equations connecting these subsystems. They have distinct XC-functional for electrons,
ions and for electron-ion interactions. The ion-ion XC functional
does not have an exchange part as the ions are treated classically, and 
the correlation part is expressed as a sum of hyper-netted-chain diagrams.
The CHNC also has three  equations, where the equations for the density
 distributions are integral equations of the HNC type. The effective classical
 temperatures of these equations should be selected to recover the actual XC-energies
 of each subsystem.

Thus the he classical map for electron-ion systems requires, in addition to the
map for e-e interactions, a map defining $\beta_{ei}V^{ei}_{cl}$ where
$\beta_{ei}=1/T^{ei}_{cls}$. Bradow {{it et al.}~\cite{Bredow15} used a diffraction
corrected Coulomb potential $V^{ei}_{cls}(r)= -|e|Z\{1-\exp(k_{ei}r)\}/r,\; |e|=1$ and an
effective temperature  $T^{ei}_{cls}=(T^{ee}_{cls}T_i)^{1/2}$. This point-ion
interaction uses the effective charge $Z$ of the ion; while the model works
successfully in some regimes, it was inaccurate, for example for aluminum and other ions
where a finite-core radius becomes necessary.

\section{Classical map applied to deuterium.}
However, the classical map applied to hydrogen or deuterium should work successfully in
regimes where the system is fully ionized, without having to worry about
pseudopotential forms. It should also work successfully even in the regime
where there are transient H-H bonds since the attractive potential arises
mostly via the exchange interaction which splits the singlet (bonding) and
triplet (anti-bonding) interactions. We associate the ``softening turnover'' (STO)
of the deuterium  Hugoniot between a density $\rho$ of 0.6 g/cm$^3$ to 0.8
g/cm$^3$ and pressures of 0.2 Mbar to 0.6 Mbar, to the changing influence of
transient D-D bonding over this interval of density and pressure
(see Fig.~\ref{hugon.fig}). In
this regime there are no stable (i.e., persistent) H-H or D-D molecules.
This  regime is
extremely sensitive to the choice of XC-functionals, finite-$T$ effects etc.,
 used in DFT calculations. The same
uncertainty reappears in the determination of the effective classical
temperature $T^{ei}_{cls}$; this is abbreviated  to $T^{ei}$ etc. in the
following. In fact, trying out different simple models for $T^{ei}$ gives
insight into the nature of the STO. 

\begin{figure}[t]
\includegraphics[width=\columnwidth]{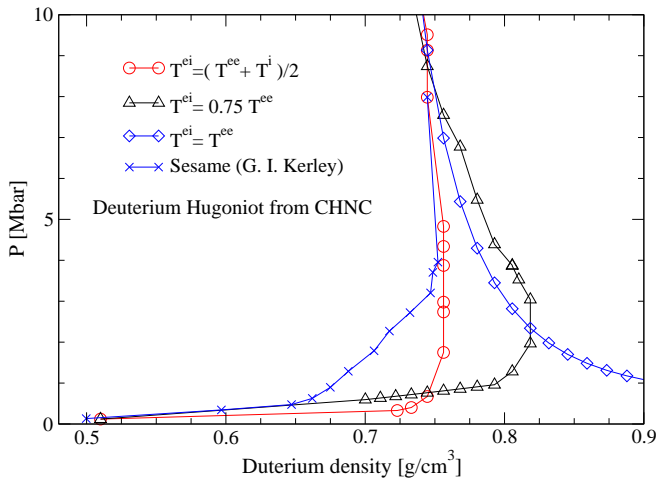}
\caption{(Color online) The deuterium hugoniots calculated using
a classical map for electrons~\cite{chnc1} and three different
assignments for the electron-ion effective temperature $T^{ei}$. This controls
the XC-effects and kinetic energy functional of electron-ion pairs.
The effective classical temperature $T^{ee}$ (energy units, Hartrees) is
 always higher than the physical temperature $T$, and becomes $\sim E_F$
as $T\to 0$. The ion temperature $T_i=T$. The three curves show that the
soft turn over (STO) region is very sensitive to the XC- and kinetic energy
functionals used. One version of the Sesame Hugoniot~\cite{Kerley83} is included
in the figure purely for comparison.} 
\label{hugon.fig}
\end{figure}

Here we consider three models of $T^{ei}$ and the corresponding CHNC-Deuterium
Hugoniots, one of which was published by Dharma-wardana and Perrot in
2002~\cite{hug02}. There we studied the choice  $T^{ei}=(T^{ee}_{cls}+T_i)/2$
(for  details see Ref.~\cite{hug02}). A more extended plot of the Hugoniot is
given in Fig.~\ref{hugon.fig}.
The free energy $F(r_s ,T)$ obtained by coupling-constant integrations over the
PDFs obtained from the CHNC method  is used to calculate the deute-
rium Hugoniot. The initial state of internal energy, volume and pressure
 $(E_0 ,V_0 , P_0 )$, with $T$=
619.6 K, initial density of 0.171 g/cm$^3$, 
 $E_0=15.886$ eV per atom, and $P_0$ = 0.0 were used. The results are
shown in Fig.~\ref{hugon.fig}.

An alternative CHNC model, with $T^{ei}=
(T^{ee}T_i)^{1/2}$ is given in Bredow {\it et al.}~\cite{Bredow15} but no
Hugoniot was calculated. That model, based on ensuring the appropriate
behaviour of the structure factors $S(k)$ as $k\to 0$~\cite{BonarthPriv15}
 may need rectification in
the limit  $T_i\to 0$. 

Since the momentum exchange in an electron-ion collision is
determined by the reduced mass $m^{ei}$ of the colliding pair, one may consider
determining $T^{ei}$ via $m^{ei}/T^{ei} = m^{ee}/T^{ee}+m^{ii}/T_i$.  This implies that
$T^{ei}=T^{ee}$ since the effective mass is determined predominantly by
the light particle, with $m^{ei}\simeq m_e$. The
corresponding Hugoniot is also presented in Fig.~\ref{hugon.fig}. This is
clearly incorrect and diverges to high compressibility 
at the onset of the STO region. In fact,
 a proper model of $T^{ei}_{cls}$ has to be based on
considerations of the electron-ion XC and kinetic  components and not on
collision dynamics which dominate high-energy collisions. A model for which
we do not yet have a Hugoniot calculation is where the classical electron-ion
interaction $\beta_{ei}V^{ei}_{cls}$ is obtained directly by an HNC
inversion of the NPA charge-density pileup at the proton.  Such a more
complete map will be considered in future work for the provision of fast, accurate
calculations of hugoniots.

\section{Conclusion}
An accurate classical representation of electrons~\cite{chnc1} is not sufficient
to deal with electron-ion systems, since the quantum features contained
in the electron-ion interaction delicately depend on including additional
quantum effects through an electron-ion exchange-correlation functional
which contributes to the e-i interaction and the classical temperature
associated with it.
The soft feature in the hugoniot is shown to be closely related to 
temperature-dependent exchange correlation effects and the onset of
 transient deuterium-deuterium molecular bonding that manifests as the
 temperatures and densities are lowered. The transient bonding is signaled
 by the appearance of pre-peaks in the
deuterium-deuterium pair-distribution functions.
Future work will present a more accurate classical  mapping of the electron-ion
interaction in the context of the models already presented~\cite{hug02, Bredow15}.
(Some of the contents of this paper were presented at  the DOE/NNSA Equation
of State workshop, 31-May, 2017, Rochester, USA).

\end{document}